\documentclass[twocolumn,twoside]{article}
\usepackage{graphics}
\usepackage{mathtext}
\usepackage{amsmath}
\usepackage{amssymb}
\usepackage[dvips]{graphicx}
\usepackage[cp1251]{inputenc}
\newcommand{\h}{\linebreak \hspace*{3ex}}
\newcommand{\hb}{\\ \hspace*{2ex}}

\begin{document}
\title{ON THE STABLE SPHERICALLY-SYMMETRIC CHARGED DUST
CONFIGURATIONS IN GENERAL RELATIVITY}
\author{Valentin D.\,Gladysh\\[2mm]
\begin{tabular}{l}
 Department of Physics,  Dnepropetrovsk National University\hb
 Gagarin Ave, 72, Dnepropetrovsk 49010,
 Ukraine, \hb \qquad \qquad \qquad {\em vgladush@dsu.dp.ua }
 \\[2mm]
\end{tabular}
}
\date{}
\maketitle ABSTRACT. The radial motion of a self-graviting charged
dust and stability condition of the static charged dust spheres
are considered. The stability is possible for the bound states of
the weakly charged layer with abnormal
charge with respect to the active mass. \\[1mm]
{\bf 1. Introduction}\\[1mm]
The collapse of a spherically-symmetric charged dust cloud is
the important problem of a General Relativity. In the papers
of Vickers P.A. (1973), Markov M.A. and Frolov V.P.,
(1970,1972), Bailyn M. and Eimerl D. (1972), Ivanenko D.D.,
Krechet V.G. and Lapchinski V.G. (1973) the solution of the
Einstein-Maxwell equations is reduced to the first integrals.
The exact solutions for charged dust spheres are obtained in
the works of Hamoui A. (1969), Bekenstein J.D. (1971), Bailyn
M., Eimerl D. (1972), Shikin I.S. (1974), Khlestkov
Yu.A.(1975), Pavlov N.V. (1976), Ori A. (1990). The main goal
for researches is to find the conditions under which the
gravitational collapse of the charged mediums is impossible.
Let us note the problem of shell crossing, which arises here.
In works of Ori A. (1991) and Goncalves S.M. (2001) it is
shown that shell crossing is inevitable in the gravitational
collapse of weakly charged dust spheres. The important problem
is also to find stability conditions of charged dust spheres.
The equilibrium  of a charged dust sphere with extremal charge
distribution was considered by Bonnor W.B (1965). The similar
problem was considered by Bonnor W.B (1993), Gladush V.D. and
Galadgyi M.V. (2007) for charged particles in the
Reissner-Nordstr\"{o}m field. In this paper we introduce the
classification of charged spherically-symmetric
configurations, and further we find the stability conditions
for them.\\[2mm]
{\bf 2. The equations of motion  for spherical layers of
the charged dust} \\
[1mm] The space-time metric has the form
\begin{equation}\label{4dds3}
  ds^{2}= \gamma_{ab}dx^{a}dx^{b} - R^2 d\sigma^2 \,,
\end{equation}
where $d\sigma^2 = d\theta^2 +\sin^2 \theta d\alpha^2 $, $
\gamma _ {ab} = \gamma _ {ab} (x^a) $ and $R=R (x^a) $, $x ^
{a} $ - time-radial coordinates $ (a, b=0,1) $.

For dynamics description of a charged dust sphere we shall
consider small region without not shell crossing. The evolution of
a spherical layers of Lagrangian radius $r \ (dr/ds=0) $, which
bounds of the sphere of the total mass $M_{tot}(r) $ and charge
$Q(r)$, can be described by the equation (see, for example, Ori A.
(1991)).
\begin{eqnarray}\label{evol}
\begin{split}
   &\left(\rho c^2\frac{d R}{ds}\right)^2= -U\equiv
   \varepsilon^2_{tot}-\rho^2 c^4 +  {}\\
   &+(\kappa\rho^2 c^2M_{tot}-\rho_{e}\varepsilon_{tot}Q)\frac{2}{R} -
   (\kappa\rho^2-\rho_{e}^2)\frac{Q^2}{R^2}\,.
\end{split}
\end{eqnarray}
Here $U=U (R, \, M_{tot}, \, Q) $ is an ``effective velocity
potential'', $ \rho $ and $ \rho_{e} $ are the dust and charge
densities, $ \varepsilon_{tot} $ is the energy density. In this
case the following relations take place:
\begin{eqnarray*}\label{are-rc2}
  \alpha(r)=\frac{\rho_{e}}{\rho c^2}
    =\frac{d Q(r)}{c^2 d{\cal M}(r)}\,, \quad
  \mathcal{H}(r)=\frac{\varepsilon_{tot}}{\rho c^2}
    =\frac{dM_{tot}(r)}{d{\cal M}(r)}\,, \quad \\
 \qquad
    \varepsilon_{u}=\frac{\varepsilon_{tot}}{\rho c^2}
    -\frac{\rho_{e}}{\rho c^2}\frac{Q(r)}{R}=
    \mathcal{H}(r)-\alpha(r)\frac{Q(r)}{R}\,,
    \qquad\qquad
\end{eqnarray*}
where $ {\cal M}(r) $ is a total rest-mass of a dust.
\\[2mm]
{\bf 3. Classification of charged dust spherically-symmetric
configurations}\\[1mm]
The character of evolution of spheres is determined by the motion
of layers

The regions of the admissible motions of the layers are
defined by the inequality $U(R)\leq 0 $, the equality $U=0 $
gives turning points $R_m $. The object of our classification
is a sphere with its boundary layer. The potential $U$ here is
not suitable, as it depends on parameter of classification --
the total energy of the sphere ${\cal E}_{tot} = M_{tot}
\,c^2$. From the equation (\ref{evol}) it follows that
$M_{tot}\geq U_{M}$. We shall accept this condition as a basis
of classification. Here
\begin{eqnarray}\label{effectfU1}
\begin{split}
   U_{M} &= \frac{1}{2\kappa\rho^2c^2}\left((
   \rho^2c^4-\varepsilon^2_{tot})R + \right.  {}\\
   &\left.  + 2\rho_{e}\varepsilon_{tot}Q + \left(\kappa\rho^2 -\rho^2_{e}\right)
   \frac{Q^2}{R}\right) \ \
\end{split}
\end{eqnarray}
is the ``effective mass potential''. This potential depends
both on global magnitudes -- the charge $Q$ and radius $R$ of
the sphere, and on local ones --- the densities $\rho \,,
\rho_{e}$ and $\varepsilon_{tot}$. Solutions of the equation
$M_{tot} = U_{M} $ define the turning points $R_m $ of the
layer. The asymptotics of the mass potential define the
character of the layers motions. Under $R\rightarrow 0 $
we have the following cases: \\
{\bf 1.1.} $U_{M} \rightarrow +\infty, \quad
\kappa\rho^2>\rho^2_{e}$ -- 
the case for weakly charged layer; \\
{\bf 1.2.} $U_{M} \rightarrow \rho_{e}\varepsilon_{tot}Q/
\kappa\rho^2, \kappa\rho^2=\rho_{e}^2$ -- the case for the layer
with an
extremal density of charge;\\
{\bf 1.3.} $U_{M} \rightarrow - \infty, \kappa\rho^2<\rho^2_{e}$
-- the case for the layer with an abnormal density of charge.
These conditions define the type of behaviour of a layer depending
on its local electrical characteristics.
On the other hand, under $R\rightarrow \infty $ we have: \\
{\bf 2.1.} $U_{M} \rightarrow +\infty, \mbox{if} \
  \rho^2 c^4>\varepsilon^2_{tot}$ -- the case for the bound states of a dust; \\
{\bf 2.2.} $U_{M} \rightarrow
    \rho_{e}\varepsilon_{tot}Q/\kappa\rho^2,  \mbox{if}
\rho^2 c^4 = \varepsilon^2_{tot}$ -- the case for the critical density of a dust; \\
{ \bf 2.3.} $U _ {M} \rightarrow - \infty, \mbox {if} \ \rho^2 c^4
< \varepsilon^2 _ {tot} $ -- the case for the unbound states of a
dust.

These conditions define the type of behaviour of a layer depending
on its local energy characteristics. Thus there are nine basic
types of behaviour of the mass potential $U_{M}$ which are
determined by local characteristics of a sphere. Besides, the
sphere of Lagrangian radius is characterized by integral
magnitudes -- $M_{tot}(r)$ and $Q(r) $. For the given total charge
$Q(r) $ the value of a
total mass $M_{tot}(r)$ determines three types of the sphere: \\
{\bf 3.1.} $ M_{tot}\sqrt{\kappa} > |Q|$ -- the weakly charged sphere; \\
{ \bf 3.2.} $M_{tot}\sqrt{\kappa} = |Q|$ -- the sphere with
an extremal charge; \\
{\bf 3.3} $M_{tot}\sqrt{\kappa} < |Q|$ -- the sphere with an
abnormal charge.

As a result the charged layers have 27 variants of motion. In the
dimensionless coordinates $\{V_{M}=U_{M}\sqrt{\kappa}/Q,\
x=c^{2}R/Q\sqrt{\kappa}\}$, these cases are illustrated at Fig.
1-9. The regions of the admissible motions are determined by
segments of an axis $x$, for which the line $V_{M}=M_{Q}$ lays
above the curve $V_{M} =V_{M}(x)$. The turning points $x_{m} \
(R_{m})$ can be found as an abscissa of intersection points of the
curve $V_{M} =V_{M}(x)$ and of the line $V_{M}=M_{Q}$. The segment
of the dashed line $V_{M}=1$ corresponds to the motion of a layer
in the field of the sphere with the extremal charge. The segments
of the dotted lines with $V_{M}>1$ and $V_{M}<1$ correspond to the
motions of the layers in a field of the weakly and abnormally
charged spheres, accordingly.
\\[2mm]
\begin{figure}[p]
\begin{center}
\includegraphics[width=7.8cm,height=6cm]{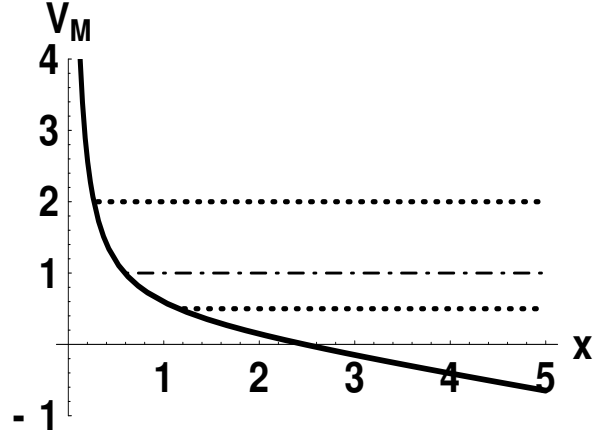}
\end{center}
\caption{The unbound states of the weakly charged layer:
$\kappa\rho^2>\rho^2_{e}\,,\ \rho^2 c^4<\varepsilon^2_{tot}$.}
\end{figure}
\begin{figure}[p]
\begin{center}
\includegraphics[width=7.8cm,height=6cm]{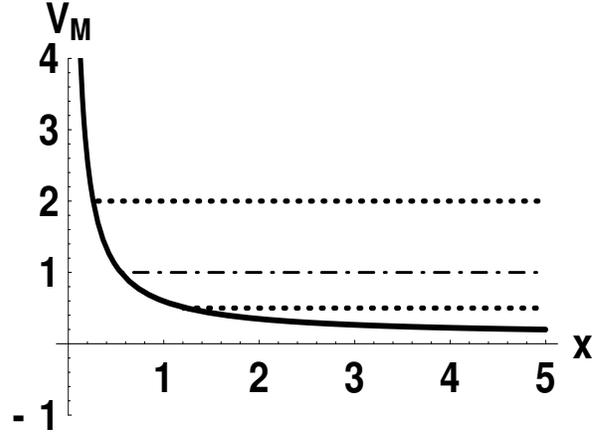}
\end{center}
\caption{The weakly charged layer with the critical density of
the dust: $\kappa\rho^2>\rho^2_{e}\,,\ \rho^2
c^4=\varepsilon^2_{tot}$.}
\end{figure}
\begin{figure}[p]
\begin{center}
\includegraphics[width=7.8cm,height=6cm]{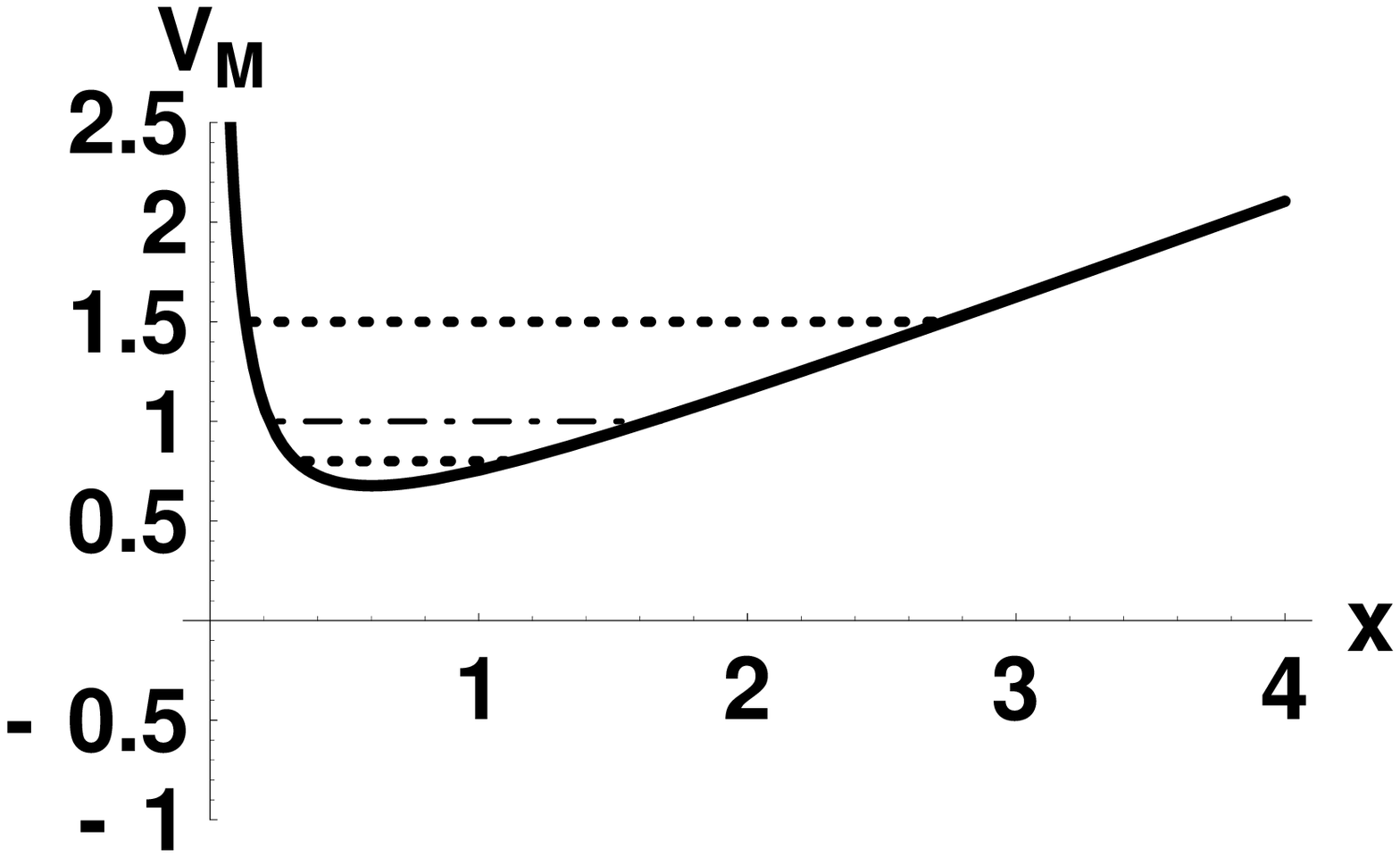}
\end{center}
\caption{The bound states of the weakly charged layer:
$\kappa\rho^2>\rho^2_{e}\,,\ \rho^2 c^4>\varepsilon^2_{tot}$.}
\end{figure}
\begin{figure}[p]
\begin{center}
\includegraphics[width=7.8cm,height=6cm]{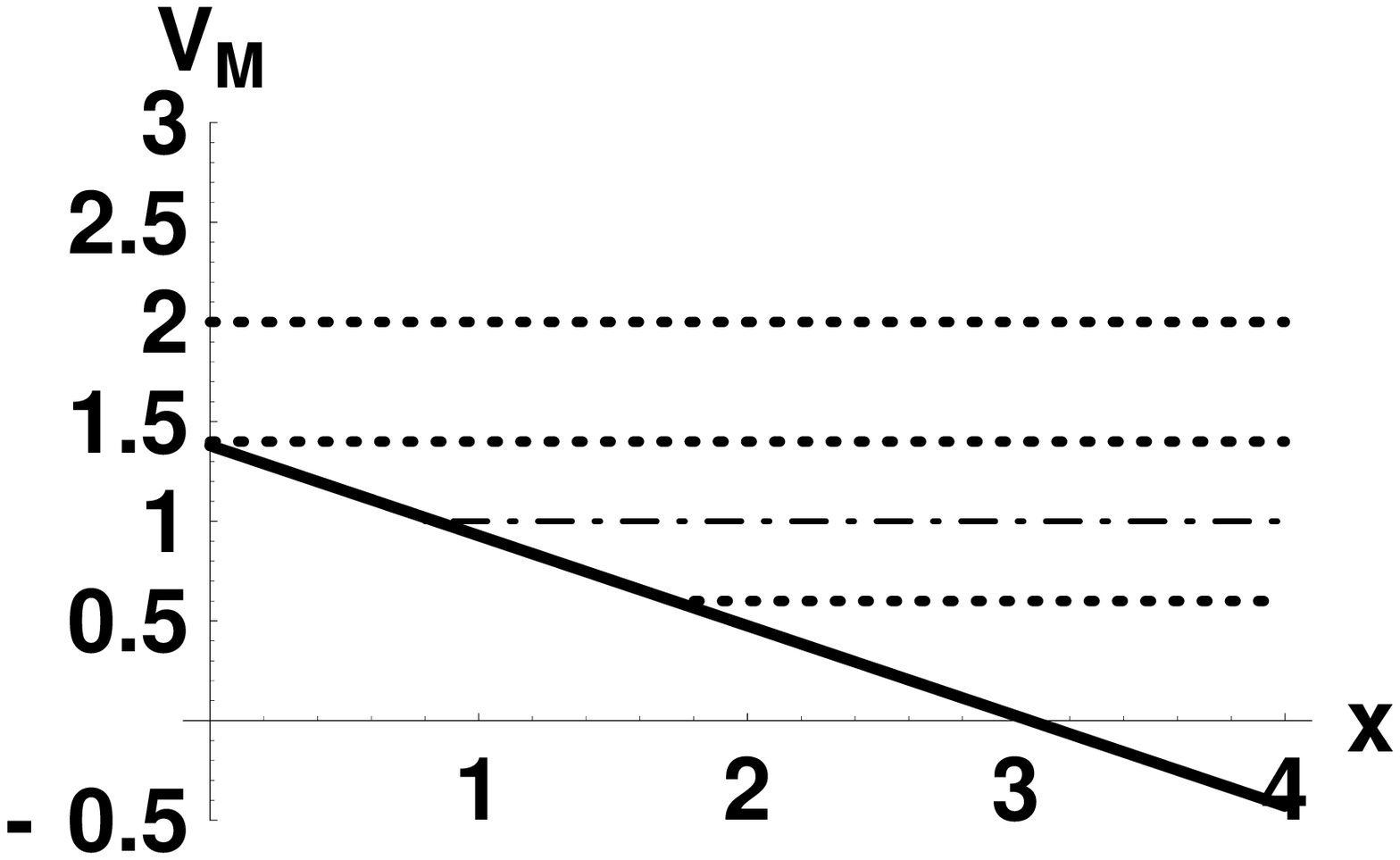}
\end{center}
\caption{The unbound states of the layer with extremal density of
the charge: $\kappa\rho^2=\rho^2_{e}\,,\ \rho^2 c^4
<\varepsilon^2_{tot}$.}
\end{figure}
\begin{figure}[p]
\begin{center}
\includegraphics[width=7.8cm,height=6cm]{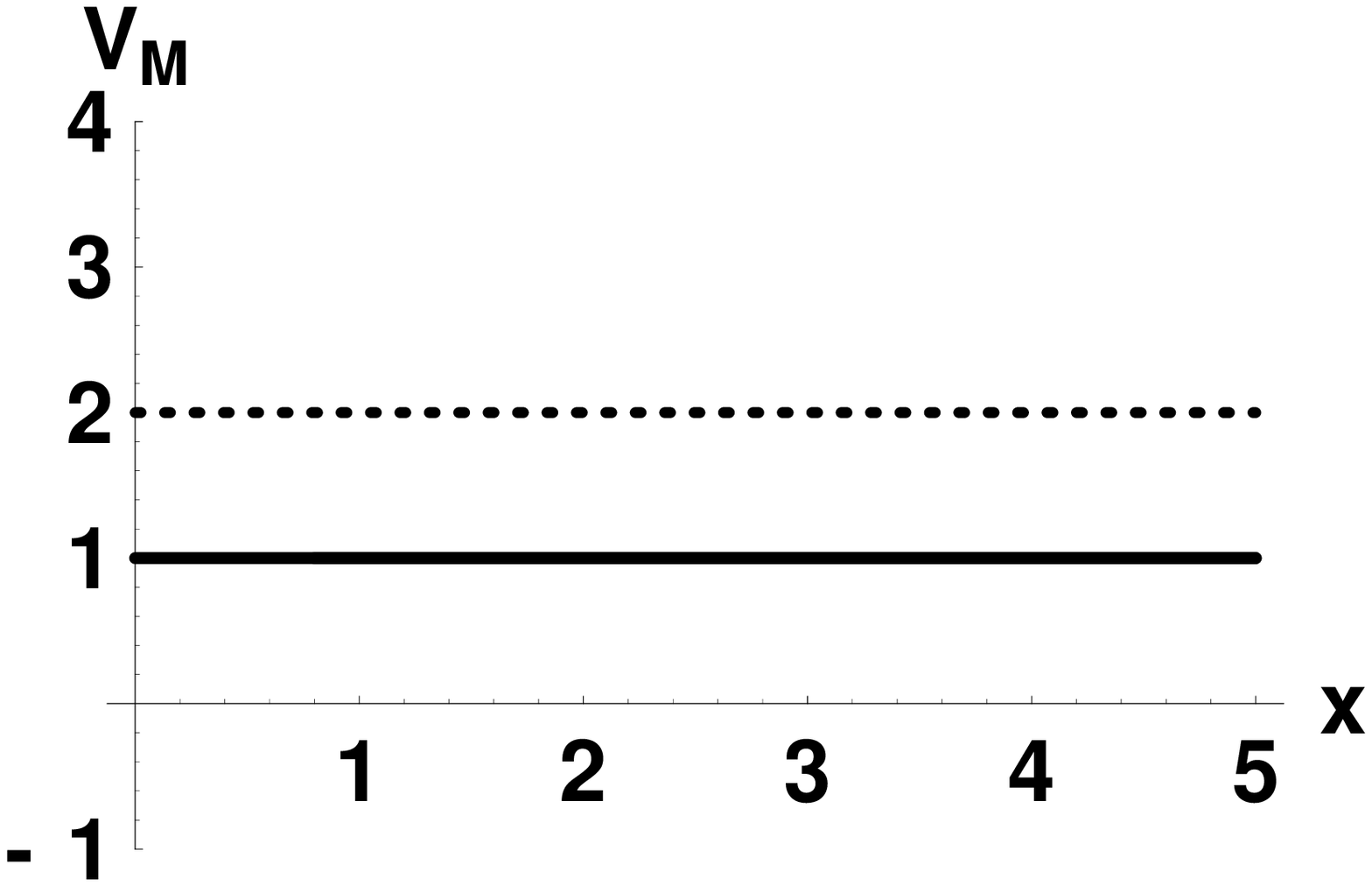}
\end{center}
\caption{The layer with extremal density of the charge and
critical density of the dust: $\kappa\rho^2=\rho^2_{e}\,,\
\rho^2 c^4 =\varepsilon^2_{tot}$.}
\end{figure}
\begin{figure}[p]
\begin{center}
\includegraphics[width=7.8cm,height=6cm]{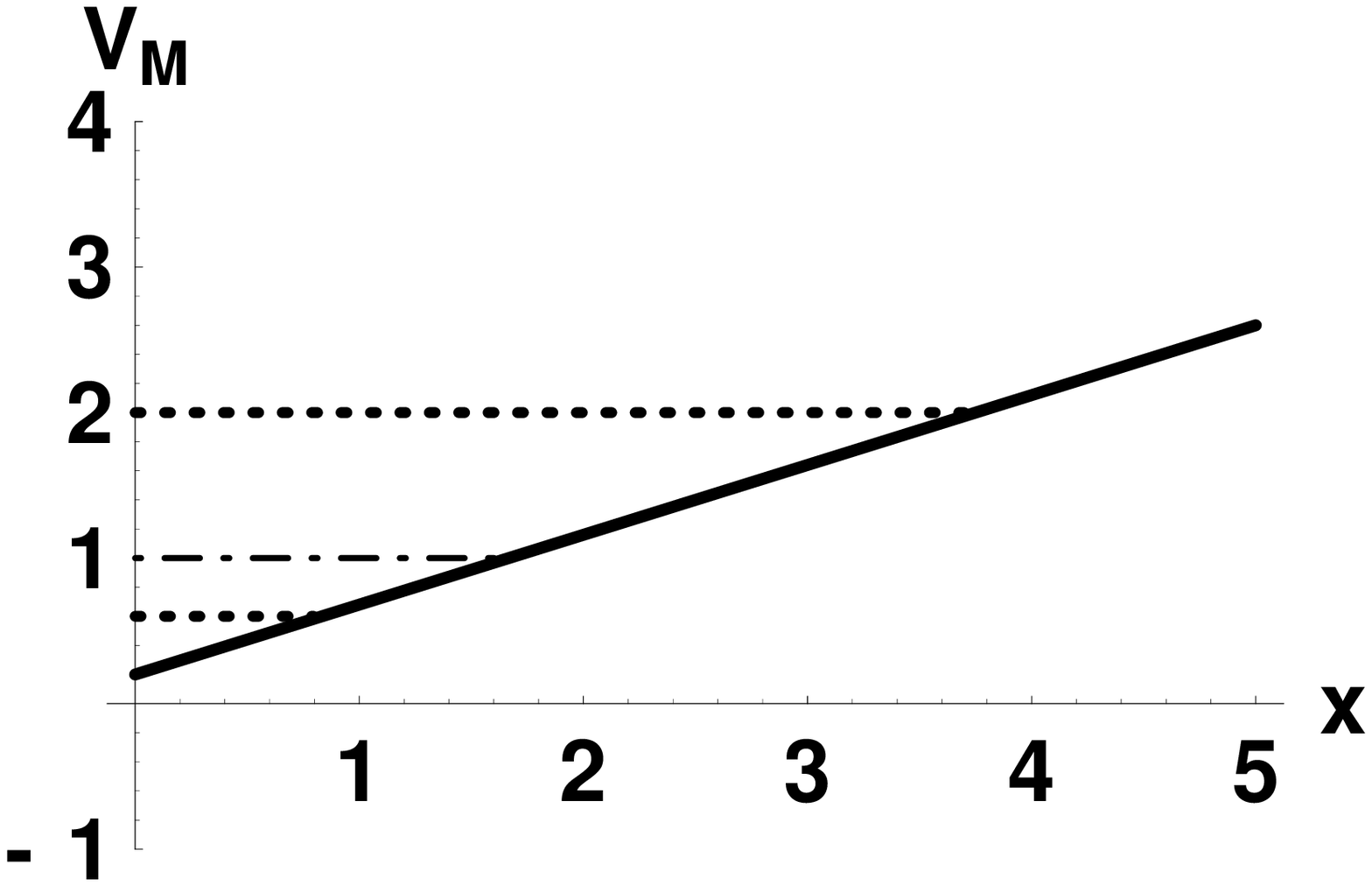}
\end{center}
\caption{The bound states of the layer with extremal charge
density: $\kappa\rho^2=\rho^2_{e}\,,\ \rho^2 c^4
> \varepsilon^2_{tot}$.}
\end{figure}
\begin{figure}[p]
\begin{center}
\includegraphics[width=7.8cm,height=6cm]{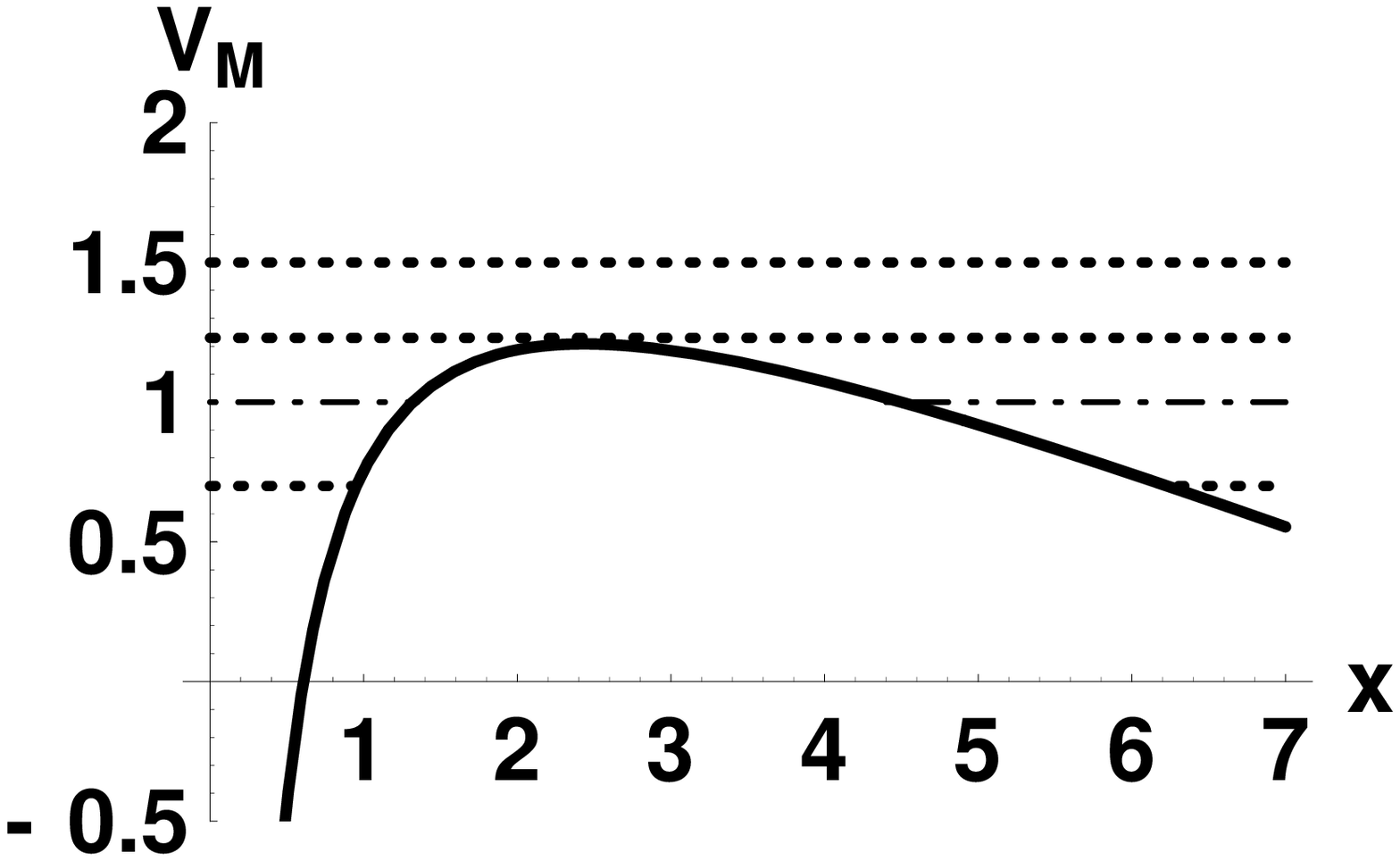}
\end{center}
\caption{The unbound states of the layer with the abnormal density
of charge: $\kappa\rho^2<\rho^2_{e}\,,\ \rho^2 c^4 <
\varepsilon^2_{tot}$.}
\end{figure}
\begin{figure}[p]
\begin{center}
\includegraphics[width=7.8cm,height=6cm]{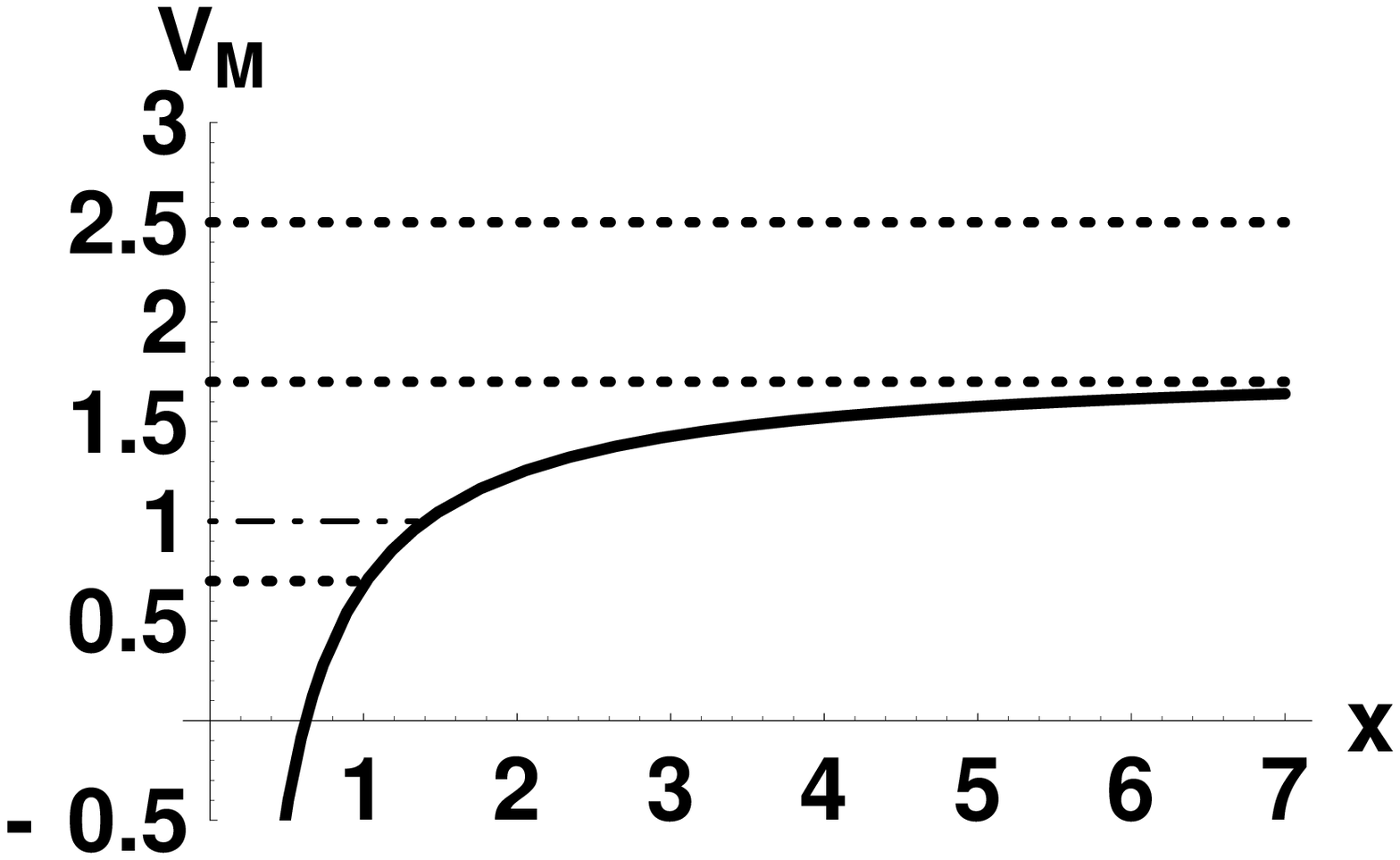}
\end{center}
\caption{The layer with the abnormal density of the charge and
critical density of the dust: $\kappa\rho^2<\rho^2_{e}$,
$\rho^2 c^4 = \varepsilon^2_{tot}$.}
\end{figure}
\begin{figure}[p] 
\begin{center}
\includegraphics[width=7.8cm,height=6cm]{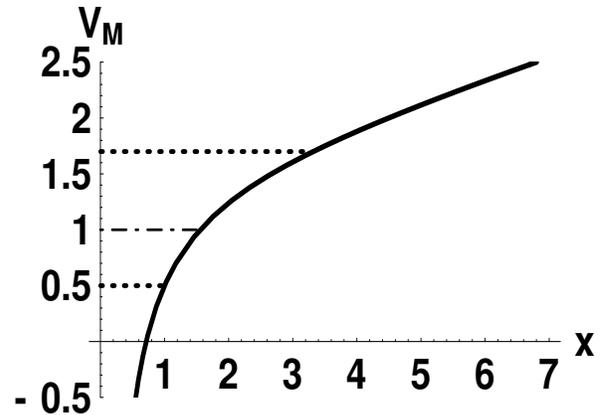}
\end{center}
\caption{The bound states of the layer with the \newline
abnormal charge density: $\kappa\rho^2<\rho^2_{e}\,,\ \rho^2
c^4
> \varepsilon^2_{tot}$. }
\end{figure}
{\bf 4. The stability conditions for spherically-symmetric
configurations of the charged dust}\\[1mm]
The stable static state of the layer is possible for bound states
of the weakly charged layer (fig. 3), when $\kappa\rho^2 >
\rho^2_{e}\,, \ \rho^2c^4>\varepsilon^2_{tot}$. If the layer is at
the bottom of the potential well the conditions of the
stationarity and equilibrium are satisfied:
\begin{eqnarray}
\begin{split}
  &\left(\frac{d R}{ds}\right)^2 =
    \left(\mathcal{H}(r)-\alpha(r)\frac{Q(r)}{R}\right)^2 - {}\\
   &-1+\frac{2\kappa M_{tot}(r)}{c^2 R}-\frac{\kappa Q^2(r)}{c^4
   R^2}=0,\qquad \label{fibermotion}
\end{split}\\
\begin{split}
 &\frac{d^2R}{ds^2} =   \left(\mathcal{H}(r)-
  \frac{\alpha(r)Q(r)}{R}\right)\frac{\alpha(r)Q(r)}{R^2} - {}\\
 &- \frac{\kappa M_{tot}(r)}{c^2 R^2}+\frac{\kappa Q^2(r)}{c^4 R^3}
 = 0. \label{fibermotion1}
\end{split}
\end{eqnarray}
Hence we obtain the following equilibrium condition for the layers
\begin{eqnarray}\label{condstat}
\begin{split}
  &\left(\frac{\kappa M_{tot}(r)}{Q(r)}-\alpha
  c^2\mathcal{H}(r)\right)^2= {}\\
 & = \left(1-\mathcal{H}^2(r)\right)
  \left(\kappa-\alpha^2c^4\right)\,.
\end{split}
\end{eqnarray}
Thus for radius of an equilibrium layer we have
\begin{eqnarray}\label{Rm130}
    R=
     \frac{\kappa M_{tot} (d{\cal M})^2 - Q d Q dM_{tot}}
    {c^2((d{\cal M})^2 -(dM_{tot})^2 )}\,,\qquad
\end{eqnarray}
that corresponds to the minimum of potential $ M_{tot}$. Using
integral magnitudes we obtain necessary and sufficient conditions
of a stable equilibrium of the charged dust configuration:
\begin{eqnarray}
   |d{\cal M}|>|dM_{tot}| \,, \quad
   \sqrt{\kappa}|d{\cal M}|>|d Q|\,,
   \label{nonequiv1}\\
   \kappa\Phi(dM_{tot}\,,d Q)\equiv
   \kappa Q^2(dM_{tot})^2 - \nonumber\\ - 2\kappa Q M_{tot}d
   QdM_{tot} + Q^2(d Q)^2= \\
 =\kappa\left( Q^2 -\kappa M_{tot}^2\right)(d{\cal M})^2 \,.
  \nonumber\qquad \qquad \label{condstatg}
\end{eqnarray}
Than it follows that
\begin{eqnarray}\label{0condstatg}
\begin{split}
   & \kappa\left(QdM_{tot}- M_{tot}d Q\right)^2 = {}\\
   & = (Q^2 -\kappa M_{tot}^2)\,
   \left(\kappa(d{\cal M})^2-(d Q)^2\right) >0 \,.
\end{split}
\end{eqnarray}
In virtue of (\ref{nonequiv1}) one can formulate the following
theorem: the stable static states are possible only for the bound
states of the weakly charged layer with the abnormal charge $Q $
with respect to the active mass $M_{tot}$:
\begin{equation}\label{condstatMtQ}
   |d{\cal M}|>|dM_{tot}| \,, \ \sqrt{\kappa}|d{\cal M}|>|d
Q|\,, \ Q^2 >\kappa M_{tot}^2\,.
\end{equation}
It follows from here, that for the stable static sphere of the
charged dust the quadratic form $\Phi(dM_{tot},d Q)$  in
(\ref{condstatg}) is positively defined $\Phi(dM_{tot},d Q)>0$.

From (\ref{0condstatg}) we  obtain one more relation for
stable static states of the charged dust sphere:
\begin{eqnarray}\label{2condstatg}
\begin{split}
   & \ \frac{\sqrt{\kappa} \,dM_{tot}}{d Q}
   = \frac{\sqrt{\kappa}\, M_{tot}}{ Q} \pm {}\\
   &\pm\sqrt{1 -\frac{\kappa M_{tot}^2}{Q^2}\ }\,
   \sqrt{\frac{\kappa(d{\cal M})^2}{(d Q)^2}-1} \,.
\end{split}
\end{eqnarray}
\newpage \noindent
This equation is invariant with respect to scale
transformation
\begin{equation}\label{podobie}
    M_{tot}(r)=a M'_{tot}(r)\,,\ Q(r) =a Q'(r)\,, \
    {\cal M}(r)=a {\cal M}'(r).\nonumber
\end{equation}
Here $R=aR'$. Thus we have a scaling law. If we multiply the
distribution functions of the charge, total and proper mass by
factor of $a$ the new configuration remains stable, and its
radius will grow by the same factor $a$.

As an example, let us consider the particle-like configuration
with parameters of an electron (mass $ m_{e}\, = \, 9,109\cdot
10^{-28}\,\mbox{\bf g}$, charge $e \, =\, 4,803\cdot
10^{-10}\,\mbox{\bf cm}^{3/2}\,\mbox{\bf g}^{1/2}\,\mbox{\bf
s}^{-1}$) and with the geometrodynamical charge
$q_{m_{e}}=\sqrt{\kappa} m_{e} = 23,53\cdot 10^{-32}\mbox{\bf
cm}^{3/2}\mbox{\bf g}^{1/2}\mbox{\bf s}^{-1} $. Since
$e/q_{me}=2\cdot 10^{21}\gg 1$, we deal here with the abnormally
charge object. Let us consider now point object with parameters of
an electron and the exterior gravitational field of the
Reissner-Nordstr\"{o}m. In virtue of inequality $e>\sqrt{\kappa}
m_{e}$ a naked singularity takes place. It contradicts to the
cosmic censorship hypothesis of Penrose (1969), according to which
the singularity should be hidden by horizon. Thus it follows, that
the electron can not be the point object. If we neglect an
intrinsic moment, the simplest classical model of such object can
be particle-like spherical configuration of the charged dust with
the total mass $M_{tot}=m_{e}$ and charge $Q=e$, which satisfies
the indicated stability conditions.\\[2mm]
\unitlength=1in \indent
{\bf References\\[2mm]}
Vickers P.A.: 1973, {\it Ann. Inst. H. Poincare,} {\bf 18},
137. \\
Markov M.A., Frolov V.P.: 1970, {\it Theor and Math. \h
Phys.,} {\bf 3}, 3.\\
Markov M.A., Frolov V.P.: 1972, {\it Theor and Math. \h
Phys.,} {\bf 13}, 41.\\
Bailyn M., Eimerl D.: 1972, {\it Phys. Rev.~D,} {\bf 5},
1897.\\
Ivanenko D.D., Krechet V.G., Lapchinskiy V.G.: \h 1973, {\it
Izv. Vyssh. Uchebn. Zaved. SSSR Fiz, } {\bf 12}, \h 63.\\
Hamoui A.: 1969, {\it Ann. Inst. H. Poincare,} {\bf 10},
195.\\
Bekenstein J.D.: 1971, {\it Phys. Rev.~D,} {\bf 4}, 2185.\\
Bailyn M., Eimerl D.: 1972, {\it Phys. Rev.~D,} {\bf 5}, 1897.\\
Shikin I.S.1: 1974, {\it JETP,} {\bf 67}, 433.\\
Khlestkov Yu.A.: 1975, {\it JETP,} {\bf 68}, 387 \\
Pavlov N.V.: 1976, {\it Izv. Vyssh. Uchebn. Zaved. SSSR \h
 Fiz,} {\bf 4}, 107.\\
Ori A.: 1990, {\it Class. Quantum Grav.,} {\bf 7}, 985.\\
Ori A.: 1991, {\it Phys. Rev.~D,} {\bf 44}, 2278.\\
Goncalves S.M.: (2001), {\it Phys. Rev.~D,} {\bf 63}, 124017.\\
Bonnor W.B.: (1993), {\it Class. Quantum Grav.,} {\bf 10}, \h
 2077.\\
Bonnor W.B.: (1965), {\it  Mon. Not. Roy. Astron. \h
Soc.}{\bf 129}, 443.\\
Gladush V.D., Galadgyi M.V.: 2007, {\it Odessa Astron. \h
Publ.,}
{\bf 20}.\\
Penrose R.: 1969, {\it Riv. Nuouo Cimento,} {\bf 1}, 252.\\

\end{document}